\documentclass[pra,aps,superscriptaddress,twocolumn,notitlepage,letter,nopacs,nofootinbib,longbibliography]{revtex4-2}

\usepackage{graphicx,color,amsmath,amsfonts,enumerate,amsthm,amssymb,mathtools,enumitem,thmtools,hyperref,subfigure,mathdots,enumitem,centernot,bm,soul,bbm,stackrel,verbatim}
\usepackage[capitalise, noabbrev]{cleveref}
\usepackage{mathrsfs}
\usepackage{babel}
\usepackage{stmaryrd}
\usepackage{multirow}
\hypersetup{colorlinks=true,linkcolor=blue,citecolor=blue,urlcolor=blue}

\usepackage{fontawesome}
\usepackage{physics}

\usepackage{tikz,pgfplots}
\usetikzlibrary{quantikz}

\def\R{ \mathbb{R} }

\def\>{\rangle}
\def\<{\langle}

\newcommand{\kkhide}[1]{}

\newcommand{\rs}[1]{{\color{blue}#1}}

\definecolor{ppblue}{RGB}{46,117,182}
\definecolor{ppred}{RGB}{197, 90, 17}

\theoremstyle{plain}

\theoremstyle{definition}

\newcommand{\robertocorr}[2]{{\color{red}\st{#2}} {\color{blue} #1}}
\newcommand{\jakub}[1]{{\color{red}({\tt JC}: #1)}}

\newcommand{\ax}[2]{ \textcolor{magenta}{({\tt AX:\ #1}) #2}}

\begin{document}

\title{Quantum Random Access Codes Implementation for Resource Allocation and Coexistence with Classical Telecommunication}
\author{Domenico Ribezzo}
\affiliation{Istituto Nazionale di Ottica del Consiglio Nazionale delle Ricerche (CNR-INO), 50125 Firenze, Italy}
\affiliation{Università degli Studi di Napoli Federico II, Napoli, Italy}

\author{Roberto Salazar}
\email{roberto.salazar@uj.edu.pl}
\affiliation{Faculty of Physics, Astronomy and Applied Computer Science, Jagiellonian University, 30-348 Kraków, Poland}

\author{Jakub Czartowski}
\affiliation{Faculty of Physics, Astronomy and Applied Computer Science, Jagiellonian University, 30-348 Kraków, Poland}
\affiliation{Doctoral School of Exact and Natural Sciences, Jagiellonian University, ul. Łojasiewicza 11, 30-348 Kraków, Poland}

\author{Flora Segur}
\affiliation{Istituto Nazionale di Ottica del Consiglio Nazionale delle Ricerche (CNR-INO), 50125 Firenze, Italy}

\author{Gianmarco Lemmi}
\affiliation{Istituto Nazionale di Ottica del Consiglio Nazionale delle Ricerche (CNR-INO), 50125 Firenze, Italy}

\author{Antoine Petitjean}
\affiliation{Istituto Nazionale di Ottica del Consiglio Nazionale delle Ricerche (CNR-INO), 50125 Firenze, Italy}

\author{Noel Farrugia}
\affiliation{Merqury Cybersecurity Limited, Malta}

\author{André Xuereb}
\affiliation{Department of Physics, University of Malta, Msida MSD 2080, Malta}
\affiliation{Merqury Cybersecurity Limited, Malta}

\author{Davide Bacco}
\affiliation{Department of Physics and Astronomy, University of Florence, 50019 Sesto Fiorentino, Italy}

\affiliation{QTI S.r.l.,  50125, Firenze, Italy}

\author{Alessandro Zavatta}
\email{alessandro.zavatta@ino.cnr.it}
\affiliation{Istituto Nazionale di Ottica del Consiglio Nazionale delle Ricerche (CNR-INO), 50125 Firenze, Italy}
\affiliation{QTI S.r.l.,  50125, Firenze, Italy}

\begin{abstract}
 In a world where Quantum Networks are rapidly becoming a reality, the development of the Quantum Internet is gaining increasing interest. 
Nevertheless, modern quantum networks are still in the early stages of development and have limited capacity to distribute resources among different users -- a constraint that needs to be taken into account. 
In this work we aim to investigate these constraints, using a novel setup for implementing Quantum Random Access Codes (QRACs), communication protocols known for their quantum advantage over their classical counterparts and semi-
device-independent self-testing applications. 
Our QRAC states, made for the first time using weak coherent pulses instead of entangled single photons, allow us to experimentally test our encoding and decoding strategy from the resource allocation perspective. 
Moreover, by emulating a coexistent classical communication, we test the resilience of our implementation in presence of noise. 
The achieved results represent a significant milestone both for theoretical studies of quantum resource allocation and for the implementation of quantum infrastructures capable of coexisting with regular telecommunication networks. 
\end{abstract}

\maketitle

\section{Introduction}

Quantum networks, and particularly the Quantum Internet, represent a flourishing field in the realm of quantum technologies, offering promising applications that are already within reach or just around the corner. A recently introduced hierarchy of development stages of Quantum Internet, depending on necessary resources and implemented functionalities, provides a scale against which one can compare different proposals and implementations~\cite{Wehner2018}. The early stages of this hierarchy, with the first two utilising trusted repeaters and prepare-and-measure devices, respectively, are likely to be the first to be implemented in real-world communication networks because of their minimal requirements of Quantum resources. Additionally, the above networks will be relevant in the future scenarios as proxies for those who do not possess direct access to more developed networks due to the lack of quantum resources.
The second stage of a Quantum Internet considers only prepare-and-measure quantum devices~\cite{Wehner2018}, which are the first to offer end-to-end quantum functionality. For example, it enables end-to-end QKD without the need to trust intermediary repeater nodes. Informally, this stage allows any node to prepare a one-qudit state and transmit the resulting state to any other node, which then measures it.


Our technique and experimental design were tested with a cutting-edge quantum communication task to showcase their power and relevance. We accomplished the above by introducing a new implementation of quantum random access codes (QRACs)~\cite{HongWei2012,Brunner2013,Takavoli2015,Takavoli2018}. We~selected QRACs as the test protocol motivated by their iconic status in the prepare-and-measure scenario ~\cite{EdgarLima2018,Edgar2018,Foletto2020} and their advantages within the broader context of future quantum technology ecosystems~\cite{purohit2023building,UniFund,EUQuantFlag,EUQuantStr}. For instance, QRACs play a crucial role as a semi-device-independent self-testing tool~\cite{Farkas2019,Mohan2019,Tavakoli2021,Wei2021}, a feature which could be exploited to remotely test quantum resources ~\cite{Farkas2019,Carmeli2020,Tavakoli2021,Heinosaari2022}. By implementing a semi-device-independent tests like QRACs, we can enhance the reliability of quantum resources, making protocols with distributed quantum operations a viable option for a wide range of applications. The above concept aligns with the current goals of quantum technology initiatives, such as the Quantum Flagship program in Europe, emphasizing technological independence and security ~\cite{EUQuantFlag,EUQuantStr,euroqci}. 

The telecommunications sector is one of the most
prevalent in the world's communication network and as such, integrating QRACs into standard telecommunication architectures holds the potential to accelerate access for quantum technology to emergent quantum R$\&$D companies and institutions. Indeed, QRACs could serve as a bridge between quantum and classical networks, offering a means to establish trust in quantum components within a classical infrastructure. However, current implementations of QRACs present severe limitations for integrating the telecommunication infrastructure. To date, all experimental  QRACs setups employ single-photon sources based on spontaneous parametric down-conversion (SPDC)~\cite{foletto2020experimental,anwer2020experimental,wang2019experimental,xiao2021widening}. While SPDC represents a consolidated method for creating entanglement and single photons, it has limitations in terms of generation rate and it is far from being a plug-and-play solution. Our QRAC encoding states are produced exploiting a more practical weak coherent pulses source. Furthermore, our protocol choice allowed us to exploit the quantum incompatibility of measurements in a unique and untested manner, here evaluated experimentally for the first time. Precisely, the measurements of this protocol achieved optimal performances in terms of communication and additionally allocate the incompatibility resource optimally~\cite{Salazar2021}.

Specifically, our experiment used time-bin and phase degrees of freedom to implement the QRAC protocol for one and two pulse-encoded qubits of a C-band laser. We exposed both one and two qubits to a depolarizing noise to test the efficiency of the protocols, emulating field trial conditions. 
This method allows us to examine the impact of noise on quantum communication rates and to test the reliability of quantum resource allocation theory; all the experimental results have been found in compliance with the theoretical model.

The fundamental step in establishing the scalability of our methods for multiple qubits is the comparison between the QRAC protocols for one and two qubits. Furthermore, our tests, which emulate standard conditions in telecommunications systems, offer strong evidence for the applicability of our technique. By this virtue, the development of early quantum networks, particularly those with limited resources, is substantially supported by our work. Furthermore, our study promotes and encourages theoretical investigations into valuable quantum resources in prepare-and-measure scenarios, providing a reference for benchmarking the near-future implementations of early Quantum Internet. Indeed, our experiment illustrates the critical role that prepare-and-measure protocols must play in the short-term applications of quantum telecommunications.

\section{Theoretical background}
 
\subsection{Quantum formalism}

In what follows we will be moving primarily within the context of quantum mechanics, where the states of finite dimension $d$ are defined as elements of Hilbert space, $\ket{\psi}\in\mathcal{H}_d$. 
Probabilistic nature of the theory is embodied in the Born rule, which defines the probability of measuring a state $\ket{\psi}$ in another state $\ket{\phi}$ as $\abs{\ip{\psi}{\phi}}^2$. This enforces the state normalization condition, $\abs{\ip{\psi}} = 1$, and equivalence up to a global phase, $\ket{\psi}\sim e^{i \theta}\ket{\psi}$ for any $\theta\in\R$.
More generally, one may consider probabilistic mixtures of quantum states, which lead to the formalism of the density matrices. We say that a density matrix $\rho$ is a pure state if it can be written as a projector on some state $\ket{\psi}$, $\rho = \op{\psi}$. In its most general form, a density matrix is defined as a hermitian semidefinite operator of unit trace, $\rho\geq 0,\,\rho = \rho^\dag$ and $\Tr\rho = 1$, and can be understood as probabilistic mixture of pure states, $\rho = \sum_i p_i \op{\psi_i}$.

Any measurement performed on a quantum state can be described using positive operator-valued measures~(POVM). A POVM with $n$ possible outcomes can be defined as a set of operators \mbox{$M = \qty{M(1),\hdots,M(n)}$} with $M(i) \geq 0$, which provides a resolution of identity, $\sum_{i=1}^n M(i) = \mathbbm{1}$. Probability of observing the $i$-th outcome is then defined as $p(i|\rho) = \Tr_i\qty(\rho M(i))$. 
In particular, in the context of this work we will be interested in so-called von Neumann measurements, defined by a set of projectors on an orthonormal basis of $\mathcal{H}_d$. Thus, given such a basis $\qty{\ket{\psi_i}}$ the corresponding measurement is defined as~$\qty{M_\psi(i) = \op{\psi_i}}$.

\subsection{Incompatibility of quantum measurements}


\label{sec:theory}
We define the incompatibility of two measurements $M_1$ and $M_2$ as impossibility of measuring both $M_1$ and $M_2$ without one of the measurements having effect on the outcomes of the other. Already in the early days of quantum theory the community recognized the pivotal role, played by the incompatibility in its basic form of noncommutativity between observables, expressed as $\comm{A}{B}\neq 0$, distinguishing them from their commuting classical counterparts. This property gave rise to the celebrated Schr{\"o}dinger-Robertson uncertainty relations~\cite{Robertson1929}. Following the modern formulation given in~\cite{Farkas2022} we say that two measurements $M_1$ and $M_2$ are compatible if there exists a parent measurement $G$ producing the former measurements as its marginals,

\begin{align*}
    M_1(x) & = \sum_y G(x,y), &
    M_2(y) & = \sum_x G(x,y),
\end{align*}
and are incompatible otherwise. 

Incompatibility is one of the most prominent properties separating the quantum theory from its classical counterpart, which can be seen already in its fundamental aspects. Thus, a resource theory of incompatibility provides an invaluable tool for quantifying this gap~\cite{Buscemi2020}.

Consider the set $\mathcal{S}$ of all possible pairs of measurements, within which one may identify the subset $\mathcal{F}$ of all compatible pairs, now understood to be the \textit{free set} of the theory. In order to quantify the resource content of any particular measurement one introduces so called \textit{monotones}; a \textit{faithful} monotone $\mathcal{M}$ is a function which assumes a zero value whenever a pair of measurements is in $\mathcal{F}$ and is greater than zero otherwise. Furthermore, a monotone is nonincreasing, $\mathcal{M}(o[\cdot]) \leq \mathcal{M}(\cdot)$, under operations $o$ from the set $\mathcal{O}$ of free operations, which include addition of noise and classical simulations~\cite{Paul2019}. Some monotones in the literature quantify the advantage of incompatible measurements over any compatible set in quantum state discrimination tasks~\cite{Paul2019} or advantage in quantum random access codes over their classical counterparts~\cite{Carmeli2020}.


A common practice in resource theories is to identify the maximally resourceful objects and subsequently exploit the resource from them. For the resource theory of incompatibility, the sets of von Neumann measurements defined by mutually unbiased bases (MUBs) are among the most incompatible~\cite{Designolle2019}. In particular, pairs of MUBs are the most incompatible pairs of measurements~\cite{Designolle2019}.  We say that a pair of orthonormal bases $\left\{|e_i\>\right\}_{i=0}^{d-1}$ and $\left\{|f_j\>\right\}_{j=0}^{d-1}$ in $\mathcal{H}^d$ is unbiased if the product of any pair of states taken from them satisfies
\begin{equation*}
    \abs{\<e_i|f_j\>}^2 = \frac{1}{d}.
\end{equation*}
Intuitively speaking, they are bases which are as different as possible with any outcome from one of them being equally probable for any state of the other. Furthermore, a set of two or more orthonormal bases is said to be mutually unbiased (MU) if they are pairwise unbiased. It is known that in dimension $d$ there exists no more than $d+1$ MUBs with saturation proven whenever the dimension is a power of prime number, $d = p^n$~\cite{durt2010}.


Additionally, a specific construction of a pair of MUBs constituted solely of product states is of particular interest for this work. Given that $\left\{|e_i\>\right\}$ and $\left\{|f_j\>\right\}$ are MU in dimension $d$, the sets

\begin{equation} \label{eq:multidim_MUBs}
    \begin{aligned}
        \{|E_I\> & = |e_{i_1}\>\otimes\hdots\otimes|e_{i_n}\>\}_{I=0}^{d^n-1}, \\
        \{|F_J\> & = |f_{j_1}\>\otimes\hdots\otimes|f_{j_n}\>\}_{J=0}^{d^n-1},
    \end{aligned}
\end{equation}
are MU in dimension $d^n$, with $i_1\hdots i_n$ and $j_1\hdots j_n$ being base-$d$ digits of indices $I$ and $J$, respectively. Consistently with our notation we designate $M_E$ and $M_F$ the POVMs associated with product bases (\ref{eq:multidim_MUBs}) respectively. 


As we will be mainly concerned with qubits, we take the pair of MU bases of interest as the eigenbases of the Pauli operators $\sigma_z$ and $\sigma_x$, 

\begin{equation} \label{eq:qbit_MUBs}
    \begin{aligned}
        |e_0\> & = |0\>, &
        |e_1\> & = |1\>, \\
        |f_0\> & = \frac{1}{\sqrt{2}}\left(|0\> + |1\>\right), &
        |f_1\> & = \frac{1}{\sqrt{2}}\left(|0\> - |1\>\right), 
    \end{aligned}
\end{equation}
with multiqubit bases defined as in \eqref{eq:multidim_MUBs}. For more thorough treatment of incompatibility and resource theories we refer the reader to the reviews~\cite{guhne2021, chitambar2019}.

\subsection{Measurement devices allocating incompatibility}

Allocating resources is a task that we encounter daily, where we have to decide how to distribute the use of limited resources. We face this challenge when dividing vegetables for a family dinner or budgeting for transportation and food expenses. Resource  allocation may seem intuitive, but it admits a rigorous mathematical formulation: it is the distribution of resources among a set of tasks, considering a practical criterion, which typically involves optimizing a figure of merit appropriate for the tasks in question. When the selected distribution method optimizes the figure of merit according to the criterion, it results in an optimal allocation of resources. As per the above definition, resource allocation is a research area widely applied across several fields, including operations research, economics, computing, communication networks, and ecology~\cite{all0,all1,all2,all3,all4,all5,all6,all7}. 

For example, \emph{proportional fairness} is a resource allocation criterion which demands that the distribution of resources between
two agents (or systems) $S_{1}$ and $S_{2}$ maximizes the following
figure of merit~\cite{Salazar2021}:
\begin{equation}
{\small\Phi_{\mathcal{Q}}\!\left(S_{1},S_{2}\right)\!=\!\log\left[\mathcal{Q}\!\left(S_{1},S_{2}\right)\right]\!+\! \log\left[\mathcal{Q}\!\left(S_{1}\right)\right]\!+\!\log\left[\mathcal{Q}\!\left(S_{2}\right)\right],}\label{eq:allocation0}
\end{equation}
where in this equation, $\mathcal{Q}\left(S_{1}\right)$ and $\mathcal{Q}\left(S_{2}\right)$
are quantifiers of the performance of $S_{1}$ and $S_{2}$ for the
respective resource allocations; likewise, $\mathcal{Q}\left(S_{1},S_{2}\right)$
quantifies the performance in the case that $S_{1}$ and $S_{2}$
act together. In the classical case, where $S_{1}$ and $S_{2}$ are
usually agents, the optimization considers trade-offs due to the finiteness
of resources and cooperation rules between $S_{1}$ and $S_{2}$ that
allow them to obtain a surplus when they act jointly. Proportional
fairness applies in cases where agents or systems should obtain resources
in a proportional or approximately equitable manner~\cite{all2,all3}, for example,
when assigning bandwidth to users who perform various online tasks
for a company.

However, the formal study of quantum resource allocation is very recent
\cite{Salazar2021}. In the quantum case, there are fundamental trade-offs between
the global resources of two systems and the ones for each system separately, such
as nonlocality and local contextuality~\cite{Deba2017}, while global surpluses result
from the superadditivity of specific quantum resources before the
result of cooperation~\cite{all1,all4}. In our work, the distributed resource
is the incompatibility of pairs of measurement devices  $\mathbf{M}_{E,F}\left[S_{1},S_{2}\right]=\left\{ M_{E}\left[S_{1},S_{2}\right],M_{F}\left[S_{1},S_{2}\right]\right\} $
acting on a pair of systems (or degrees of freedom) $\left[S_{1},S_{2}\right]$. Then according to reference~\cite{Salazar2021} the figure of merit (\ref{eq:allocation0}) must be modified to take into account the explicit use of the resource exploited, in particular
the figure of merit (\ref{eq:allocation0}) for incompatibility quantified
with a monotone $\mathcal{M}$ turns into:
\begin{equation}
{
    \begin{aligned}
    \Phi_{\mathcal{M}}&\left(\mathbf{M}_{E,F}\left[S_{1},S_{2}\right]\right)=\log\left[\mathcal{M}\left(\mathbf{M}_{E,F}\left[S_{1},S_{2}\right]\right)\right]\\&
    +\log\left[\mathcal{M}\left(\mathbf{M}_{E,F}\left[S_{1}\right]\right)\right] +\log\left[\mathcal{M}\left(\mathbf{M}_{E,F}\left[S_{2}\right]\right)\right],\label{eq:allocation1}
    \end{aligned}}
\end{equation}
where $M_{E}\left[S_{i}\right]$ means the reduction of the bipartite
measurement $M_{E}\left[S_{1},S_{2}\right]$ to system $S_{i}$ and
similarly for $M_{F}\left[S_{i}\right]$, and $\mathbf{M}_{E,F}\left[S_{i}\right]=\left\{ M_{E}\left[S_{i}\right],M_{F}\left[S_{i}\right]\right\} $. It is worth noting that when analyzing classical resources, the quantifiers $\mathcal{Q}$ in (\ref{eq:allocation0}) are typically obtained informally based on some intuitive understanding of the resource. However, the research in~\cite{Salazar2021}  provides a rigorous formalization of the quantifier designation through the theoretical framework of quantum resource theories, which we apply to formulate (\ref{eq:allocation1}).

In~\cite{Salazar2021}, the authors demonstrated that for a particular quantifier,
the figure of merit (\ref{eq:allocation1}) reaches its maximum with
measurements given by the product MUBs (\ref{eq:multidim_MUBs}) presented in the previous section.
Here it is relevant to remark that not all multipartite MUBs optimize Eq. 
(\ref{eq:allocation1}); for example, for MUBs composed from maximally entangled states,
their reduced measurements  $\mathbf{M}_{E,F}\left[S_{i}\right]$
are compatible measurements and therefore, the expression (\ref{eq:allocation1})
is sub-optimal. In this work, we use a different quantifier from the
one used in~\cite{Salazar2021}; however, it satisfies the same properties used
in the proof provided therein. Furthermore, the monotone quantifier used
in this work characterizes the advantage of decoding devices in the
experimentally implemented communication protocol, conferring an operational
meaning to using the resource.

\subsection{Quantum random access codes}
\label{sec:qrac}
\begin{figure}[ht] 
    \centering
    \includegraphics[width=\columnwidth]{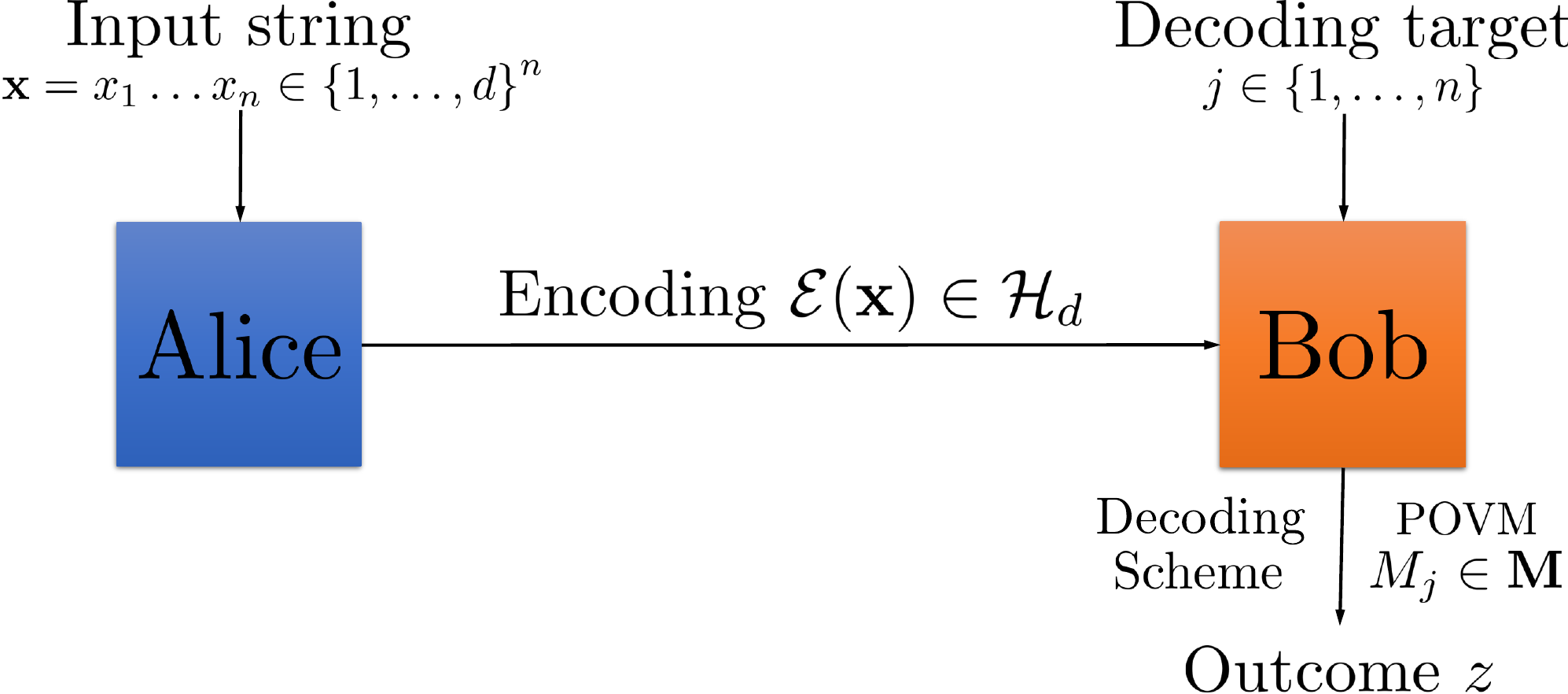}
    \caption{\emph{QRAC scheme} -- Alice receives an input string $\vb{x}$ consisting of $n$ letters from $d$-element alphabet and encodes it into a single $d$-dimensional quantum state $\mathcal{E}(\vb{x})$. Bob receives both the encoding state and a decoding target $j$, pointing to a specific letter $x_j$ to be decoded. Bob chooses his measurement $M_j\in\mathbf{M}$ and outputs a letter $z$ with probability $\Tr[M_j(z) \mathcal{E}(\vb{x})]$.}
    \label{fig:QRAC_scheme}
\end{figure}
 
Quantum random access code (QRAC) is a prepare-and-measure communication
protocol with quantum advantage which has been extensively studied
in quantum information~\cite{HongWei2012,Brunner2013,Takavoli2015,Takavoli2018,Edgar2018,Farkas2019,Carmeli2020},
including experimental applications~\cite{EdgarLima2018,Foletto2020}. In
the $(n,d)$-QRAC scheme (See Fig.~\ref{fig:QRAC_scheme}) Alice receives an input string consisting of $n$ dits, $\mathbf{x}=(x_{1},...,x_{n})$, taken from a uniform distribution, which 
she  encodes into a single qudit and sends  to Bob. In addition, Bob receives a random number
$j\in\left\{ 1,...,n\right\}$ and his task is to guess the corresponding
dit $x_{j}$. He does this by performing a measurement, depending
on $j$, thereby obtaining an outcome $z$. The encoding-decoding
is successful if \mbox{$z=x_{j}$}. 
In order to quantify the performance of the encoding-decoding scheme employed we will consider the average success probability for all possible input string $\vb{x}$ and decoding targets $j$.

The strategy of Alice and Bob consists of $d^{n}$ quantum states for
encoding and $n$ $d$-outcome measurements for decoding, all defined
for a $d$-level quantum system. We denote by $\mathcal{E}:\qty{1,\hdots,d}^n\rightarrow\mathcal{H}_{d}$ the encoding
map and $M_{1},...,M_{n}$ the measurements. Hence, $\mathcal{E}(\mathbf{x})$
is a quantum state for each $\mathbf{x}$, and $M_{1},...,M_{n}$
are $d$-outcome positive operator valued measures (POVMs). The average
success probability is then given as:
\begin{equation}
P\qty(\mathcal{E},\qty{ M_{k}} _{k=1}^{n})=\frac{1}{nd^{n}}\sum_{\mathbf{x}}\mathrm{Tr}\left[\sum_{k=1}^{n}\mathcal{E}(\mathbf{x})M_{k}\left(x_{k}\right)\right].
\end{equation}
We denote by $P_{\text{QRAC}}^{n,d}$ the best achievable average success
probability in $(n,d)$-QRAC. In the case $n=2$, it is known from~\cite{Takavoli2015}
that:
\begin{equation}
P_{\text{QRAC}}^{2,d}=\frac{1}{2}\left(1+\frac{1}{\sqrt{d}}\right).
\end{equation}
Moreover, for any dimension $d$ a pair of mutually unbiased bases together with optimal encoding saturates the above upper bound~\cite{Brunner2013,Takavoli2015,Edgar2018,Farkas2019,Carmeli2020}.
The relevance
of QRACs lies in their advantage over any classical random access
code (RAC), in which Alice sends a single dit to Bob
rather than a qudit, and Bob only has access to classical decoding schemes.
For comparison, the optimal average success probability for $(2,d)$-RACs is~\cite{Takavoli2015}:
\begin{equation}
P_{\text{RAC}}^{2,d}=\frac{1}{2}\left(1+\frac{1}{d}\right).
\end{equation}

It is evident that the probabilities satisfy the equality:
\begin{equation}
    P^{2,d^2}_{\text{QRAC}} = 
    P^{2,d}_{\text{RAC}},
\end{equation}
which translates to a second facet of the advantage of QRACs.  When the probability of successfully decoding a message is held constant, QRACs enable the transmission of twice the quantity of information bits per message compared to the conventional RAC protocol. 



Furthermore, we highlight that the benefits offered by QRACs extend beyond the aforementioned success probability enhancement, for example by providing a higher level of trust in quantum communication or computation systems \cite{Farkas2019,Mohan2019}. Indeed, QRACs lead to a quantum \emph{semi-device-independent self-testing protocol}, which is a method that allows two parties to infer the properties of a quantum state and the measurements performed on it without requiring full information about the devices used, relying on minimal assumptions and limited device knowledge \cite{Supic2020Review,Supic2020}. The integration of QRACs to the standard telecommunication infrastructure provides the first step to later integrate similar prepare-and-measure self-testing protocols, certifying additional quantum resources, for instance \cite{Tavakoli2021,Miklin2021}. 


Now, in every QRAC a given pair of decoding measurements $\mathbf{M}=\left\{ M_1,\,M_2\right\}$ has the optimal encoding if every state $\mathcal{E}(\mathbf{x})$, is the eigenvector to the maximal eigenvalue of the corresponding operator $\sum_{k=1}^{2}M_{k}\left(x_{k}\right)$~\cite{Takavoli2015,Carmeli2020}. 
In consequence we can write the maximal success probability
$P_{\text{QRAC}}\left(\mathbf{M}\right)$ using $\mathbf{M}$ as:
\begin{equation}
P_{\text{QRAC}}\left(\mathbf{M}\right)=\frac{1}{2d^{2}}\sum_{\mathbf{x}}\left\Vert \sum_{k=1}^{2}M_{k}\left(x_{k}\right)\right\Vert ,
\end{equation}
where $\left\Vert \cdot\right\Vert $ is the operator norm.
In order to quantify the allocation of the advantage over the classical random access codes, we will define the quantity

\begin{equation}
    \mathcal{A}(\mathbf{M}) = \max\qty{P_{\text{QRAC}}(\mathbf{M}) - P^{2,d}_{\text{RAC}},0},
\end{equation}
which corresponds to the excess of the average success probability above the classical bound. Since for every dimension $d$ pairs of measurements defined by MUBs, $\mathbf{M}_{ef} = \qty{M_e,\,M_f}$, saturate this quantifier up to the value:

\begin{equation}
    \mathcal{A}(\mathbf{M}_{ef}) = \frac{\sqrt{d}-1}{d},
\end{equation}
we can use a variation of Theorem 1 from~\cite{Salazar2021} in order to show that product MUBs defined in \eqref{eq:multidim_MUBs} are optimal for allocation criteria of proportional fairness. This makes them the natural measurement of choice, which we will use in order to experimentally verify the allocation of incompatibility provided by the in-lab implementations of such measurements.


\section{Methodology}
Inspired by state-of-the-art practical QKD setups, we propose a time-bin encoding scheme for the QRAC states generated through a fully in-fiber setup. More precisely, the state's production begins with a continuous wave laser, which is modulated by a cascade of intensity and phase modulators and is finally attenuated down to the single-photon level. 
A setup of this type offers significant advantages, including robustness, remote controllability, and no necessity of specific adjustments once it is installed in a rack within a telecom center. All the modulators and the optoelectronic components can operate at high-frequency regimes, matching the latest demands of the industry. Furthermore, this setup is easily reprogrammable to handle different quantum communication protocols. The same setup was used to manage both QRAC(2,2) and QRAC(2,4), and it can also be employed for quantum cryptography protocols. However, compared to real single-photon sources, this type of transmitter is affected by the phenomenon of multi-photon states. In QKD protocols, this introduces a security loophole, a problem that has been solved by introducing the decoy-state method. By randomly varying the intensity of the transmitted pulses, Alice makes it impossible for a potential eavesdropper to keep the photon statistics constant. In the case of QRAC, a multi-photon event does not introduce security issues, but it makes the protocol meaningless since its purpose is precisely to encode a certain number of bits in a single quantum object (qubit). To overcome this problem, the average number of photons per pulse has been set to a very low value. In this way, while not completely preventing the generation of multi-photon states, its probability is drastically reduced.

\section{Experimental set-up}
\begin{figure}
    \centering
    \includegraphics[width=\columnwidth]{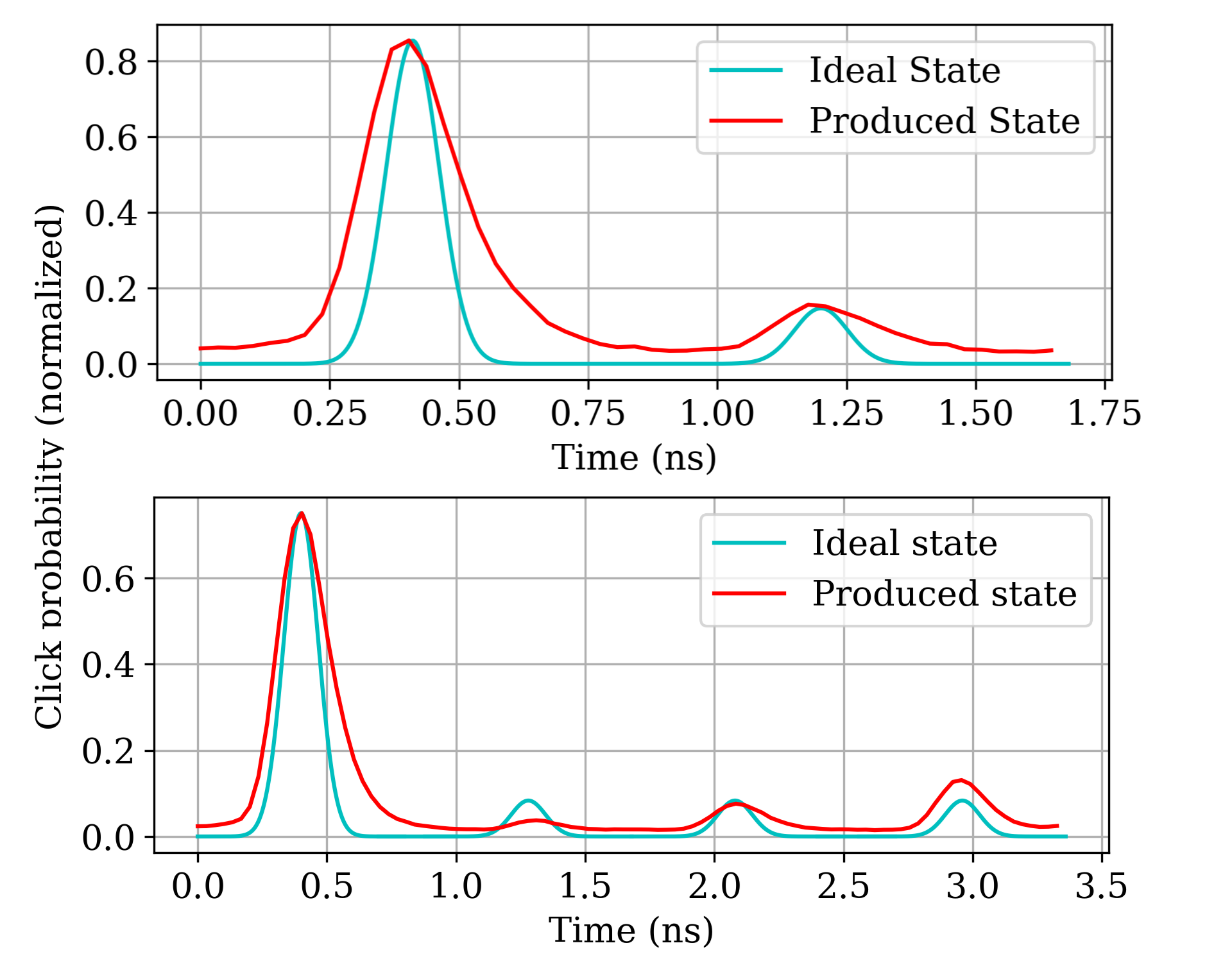}
    \caption{\textbf{QRAC(2,2) (top) and QRAC(2,4) (bottom), example of a state.} The top (bottom) figure shows the shape of the wave function of a photon encoding a QRAC(2,2) (a QRAC(2,4)). The cyan line represents the theoretical simulation, the red line is the measured profile. The two (four) pulses sizes are proportional to the values reported in~\ref{tab:enc_2d} (\ref{tab:enc_4d}), while the relative phase between the pulses can be zero or $\pi$.}
    \label{fig:statesqrac}
\end{figure}

Both the transmitter and receiver setups work in the practical C-band so that they can be fully integrated into the standard telecommunication infrastructure. 
The states, made by unbalanced pulses with a fixed relative phase, as graphically illustrated in Fig.~\ref{fig:statesqrac}, are generated by Alice carving a continuum wave C-band laser set at 1551.72 nm, corresponding to channel 32 of the International Telecommunication Union (ITU)  Dense Wavelength Division Multiplexing (DWDM) grid ~\cite{itu_dwdm}. A Field Programmable Gate Array (FPGA) board is in charge of generating all the electrical signals necessary for driving the optical components. A first intensity modulator driven by a 1.2GHz signal creates a train of 800 ps distanced pulses; a second intensity modulator, driven by a signal generated according to a pseudo-random binary sequence (PRBS), takes care of the partial suppression of some pulses according to the amplitudes given in Tab.~\ref{tab:enc_2d} and Tab.~\ref{tab:enc_4d}. In intensity modulators, the signal voltage value that operates the best extinction ratio is $V_{\pi}=12V$, but the zero is not necessarily aligned with the system ground. By adjusting the bias and the amplitude of the radio frequency signal driving the modulator, it is possible to attenuate an incoming pulse by a desired value. Subsequently, a phase modulator driven by the same PRBS inserts a pi-phase between some of the pulses. It is worth pointing out that all the modulators act on the optical signal at different moments -- the light does not propagate instantaneously! -- so, unless these delays are compensated by introducing specific delays in the radiofrequency signals, the same PRBS utilized for two modulators acts as two different sequences. The final sequence needs to be reconstructed by correlating the detected events with all the cyclic permutations of the original known sequence. Tailoring the FPGA sequences on the electric and optical delays, it is possible to avoid the last step; however, our strategy offers higher reconfigurability and, thanks to the PRBS, allows us to randomly produce all the different states in every run with a high repetition rate. The first intensity modulator produces a train of $n_p=1.190772736\cdot 10^9$ pulses per second, so the upper bound for the rate of states per second is $n_p/2$ for QRAC(2,2) and $n_p/4$ for QRAC(2,4). However, only occurrences in which the PRBS has two immediately successive different values can be used to construct a quantum state suitable in the two-dimensional QRAC protocol; concerning the four-dimensional QRAC, it is necessary to seek situations where the PRBS exhibits one high level and three low ones within a sequence of four successive values ('1000' and its permutations). In our PRBS, the first case happens 2048 times over a full cycle of 4095 bi-dimensional states, while the second case occurrences are 522 over 2047 four-dimensional states. Each cycle is repeated 145358 times per second. As a result, our generation rates are $\nu_{2D}=2048*145358=297.693184$ MHz and $\nu_{4D}=522*145358=75.876876$ MHz. In Table \ref{tab:rate_comp} we report a comparison with the quantum state generation rate of some relevant QKD implementations. In the case of the phase modulator, its effect has been initially converted into an intensity modulation in order to visualize it and compute the PRBS shift with respect to the original sequence. A PM affects just the photons with a polarization aligned to its axis, so the goal has been achieved by introducing a rotation of $\pi/4$ to the polarization of the incoming photons and then filtering out half of the photons by a polarizing beam splitter~(PBS).

In the two-dimensional protocol, the result is a train of states, each made of two pulses distanced $\tau=800$ ps, whose size is proportional to $a_1^2$ or $b_1^2$, as defined in Tab.~\ref{tab:enc_2d}, and whose relative phase is 0 or $\pi$. The optimal values $a_1$ and $b_1$ have been calculated as explained in section~\ref{sec:qrac}. Since the states are produced according to a PRBS, a sifting stage is required in post-processing to extract the states described in Tab.~\ref{tab:enc_2d}. In the four-dimensional protocol, the states are composed of four pulses. Finally, a variable optical attenuation stage is in charge of decreasing the intensity of the states down to the single-photon level. With our method, we achieved a quantum state generation rate of 298 MHz for QRAC(2,2) and 76 MHz for QRAC(2,4). A comparison with the state generation rate of QKD state-of-the-art systems is provided in Table \ref{tab:rate_comp}.
\begin{figure*}[ht]
    \centering
    \includegraphics[width=0.9\textwidth]{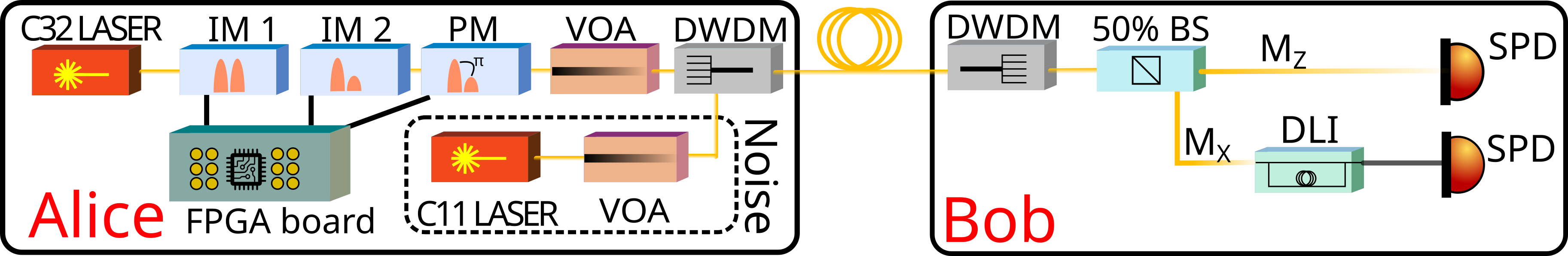}
    \caption{\textbf{Scheme of the setup.} Alice creates the quantum states with a 1551.72 nm laser (C32), whose light passes through two intensity modulators (IM1 and IM2), a phase modulator (PM) and a variable optical attenuator (VOA). A field programmable gate array (FPGA) drives the modulator. A second laser at 1568.11 nm (C11) and attenuated with a second VOA is combined with the quantum signal by a dense wavelength division multiplexing (DWDM) filter to emulate a classical transmission ongoing into the same fiber channel. Bob uses a second DWDM to demultiplexer the quantum states out of the noise, then a 50\% beam splitter (BS) routes half of the states toward a single photon detector (SPD) for the Z basis measuremente ($M_Z$), and the other half toward a delay line interferometer (DLI) to perform the X basis measurement ($M_X$).}
    \label{fig:setup}
\end{figure*}

At this point the photons are transmitted through a fiber channel exhibiting a 10 dB attenuation. At the receiver side, a 50:50 fiber beam splitter serves as a passive basis choice. For the time (\textbf{Z}) basis, the arrival time of the photons is detected using a single photon detector (SPD), an InGaAs single photon avalanche diode (SPAD) ID221 by IDQuantique, showing a dark count rate $\tau_{DC}=2500$ Hz, a detecting efficiency $\eta=20\%$ and a timing jitter $\sigma_{j}\approx 200$ ps . In the phase basis, the relative phase of the two pulses is measured by sending them into a delay line interferometer (DLI), a Mach-Zehnder interferometer with one arm longer $\tau\cdot c=28.98$ cm with respect to the other. This makes the two pulses overlap, allowing the phase measurement. 
Finally, a simultaneous classical communication is emulated by injecting a continuum wave laser with a wavelength of 1568.11 nm into the transmission channel. The classical signal is multiplexed and then filtered out by two DWDM filters; however, due to the high power of the classical signal, non-linear effects such as Raman and Brillouin scattering occur in the optical fiber channel, generating photons across a wide range of wavelengths. As a result, some of these photons cross the DWDM filter and cause noise.

\section{Results}
\subsection{QRAC(2,2)}
In QRAC(2,2), a qubit encodes a two-bit message; the couple of mutually unbiased bases utilized for encoding and then measuring the states are the bases $\textbf{Z}$ and $\textbf{X}$, composed by the eigenvectors of the Pauli matrices $\sigma_Z$ and $\sigma_X$.
In our implementation, measuring in \textbf{Z} basis corresponds to measuring the arrival time of a photon, while the \textbf{X} basis information is found by checking the relative phase between the pulses composing a state. The states are prepared according to the scheme reported in Tab.~\ref{tab:enc_2d}. These values are extracted following the rule illustrated in Sec.~\ref{sec:qrac}.

\begin{table}[h]
    \centering
    \begin{tabular}{c|c|c}
         Message & $\ket{0} \text{component}$ & $\ket{1} \text{component}$  \\
         \hline
         00 & $a_1$ & $+b_1$\\
         01 & $a_1$ & $-b_1$\\
         10 & $b_1$ & $+a_1$\\
         11 & $b_1$ & $-a_1$\\
    \end{tabular}
    \caption{Scheme of the states encoded for QRAC(2,2);\\ $a_1=\frac{\sqrt{2+\sqrt{2}}}{2}$ and $b_1=\frac{\sqrt{2-\sqrt{2}}}{2}$, the sign in front of the $\ket{1}$ component represents the relative phase between the two pulses (+ is zero and - is $\pi$).}
    \label{tab:enc_2d}
\end{table}
The states are prepared in classical regime adjusting the modulators bias according to the feedback provided by a photo-diode and a 16 GHz bandwidth oscilloscope. Once the pulses are proportional to $a_1^2$ and $b_1^2$, the mean number of photons per pulse $\mu$ is decreased to $\mu=0.2$ in order to strongly reduce the generation of multiphoton states. The probability $P_\mu(n)$ of having a number $n$ of photons contained in an independent pulse follow the Poissonian distribution:
    \begin{equation}
        P_\mu(n)=\frac{\mu^n}{n!}e^{-\mu},
    \end{equation}
    from which follows that the probability of having a multiphoton event is:
    \begin{equation}
        P_{mp}(\mu)=1-P_\mu(0)-P_\mu(1).
    \end{equation}
$P_{mp}(0.2)=0.017$ is small enough to produce a negligible error in the measurements. 

The states are measured in the Z and X bases, and the results can be found in Tab.~\ref{tab:2d_state_meas}. In QRAC(2,2), where a qubit contains the information of two bits, the Z basis measurement accesses the information of the first bit, while the X basis measurement provides the second one. What Tab.~\ref{tab:2d_state_meas} shows is the probability \textit{p} of measuring correctly the bit. Recalling that the lower threshold for \textit{p} is given by the classical RAC and is equal to $p^{CRAC(2,2)}=1/2(1+1/2)=0.75$~\cite{Takavoli2015}, the probability of obtaining a successful measurement is higher than the classical counterpart for all the generated states, i.e. there is a quantum advantage $\delta A$, where $\delta A=max\{p^{QRAC}-p^{CRAC},0\}$. 

\begin{table}[h]
    \centering
    \vspace{3mm}
    \begin{tabular}{c|c|c}
         State & $p_Z$ & $p_X$\\
         \hline
         00 & 0.8537 & 0.8502\\
         01 & 0.8532 & 0.8140 \\
         10 & 0.8520 & 0.8184\\
         11 & 0.8555 & 0.7937\\
    \end{tabular}
    \caption{$p_Z$ and $p_X$ are the probabilities \textit{p} of retrieving the correct bit in the time and in the phase basis respectively. The lower values of $p_X$ stems from heightened susceptibility of interferometric measurements to environmental noise, such as mechanical vibrations, airflows, temperature fluctuations, and more.}
    \label{tab:2d_state_meas}
\end{table}

\begin{figure}[h]
    \centering
    \includegraphics[width=0.9\columnwidth]{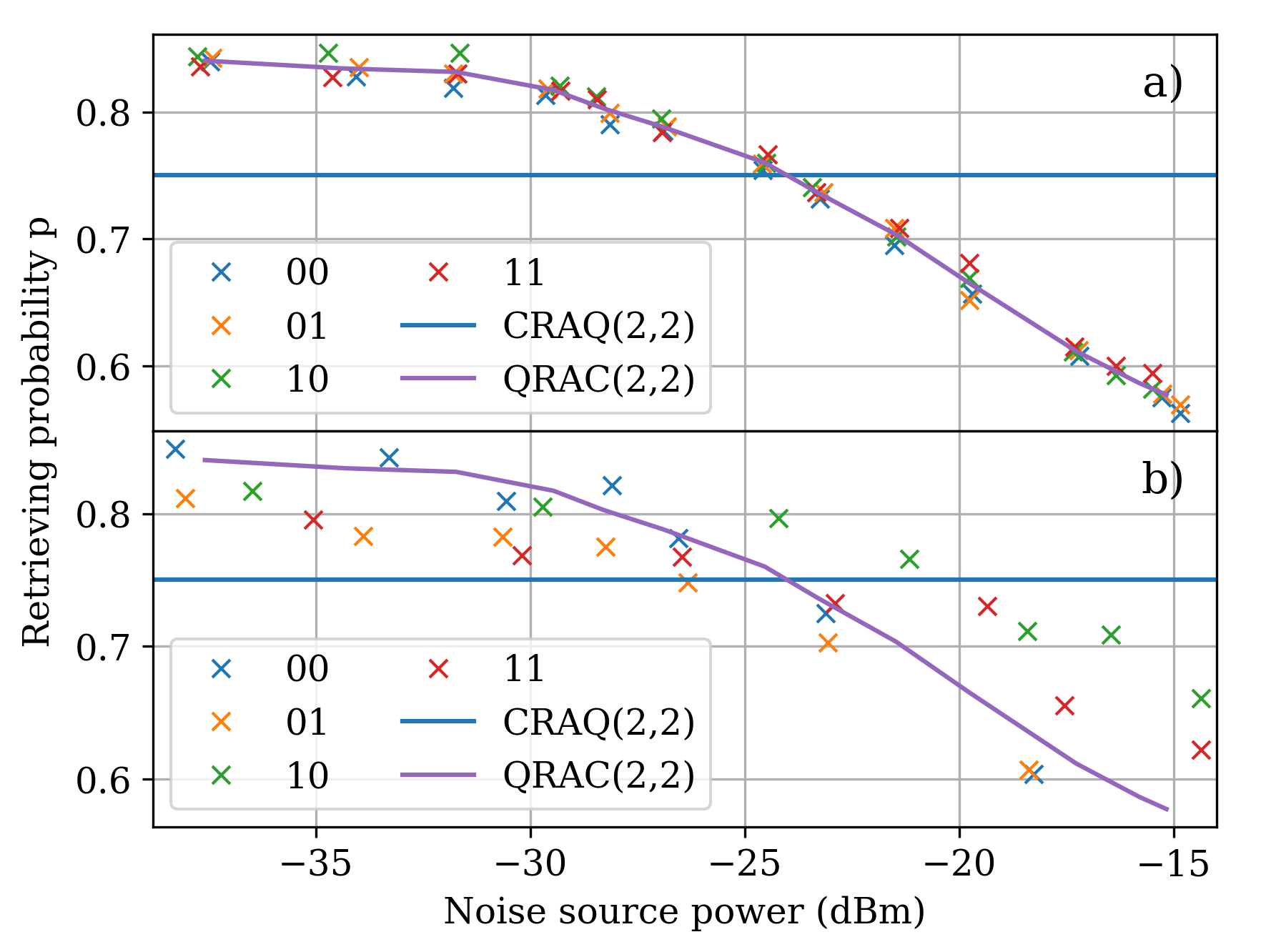}
    \caption{Probability \textit{p} of retrieving the correct value for the bit in the \textbf{Z basis (a)} and in the \textbf{X basis (b)}  versus optical power of the coexisting classical communication. The blue line represents the threshold given by the successful retrieving probability using the classical RAC protocol ($P^{CRAC}$), while the purple line shows the average probability among the different states.}
    \label{fig:QRAC_2D}
\end{figure}

Subsequently, the 1568.11 nm (channel 11 of the DWDM ITU grid) laser of the noise source is switched on and the measurement are repeated for different optical powers. For our quantum communication, this is exactly equivalent to a coexisting classical communication. Our QRAC implementation proves to be resilient for classical communication injected into the same channel with a power of up to around -25 dBm. The detailed results of the measurements in both bases are reported in Fig.~\ref{fig:QRAC_2D}. The phase measurements show bigger fluctuations compared to the time measurements, due to the higher sensitivity of an interferometric setup to noise and disturbances.

\subsection{QRAC(2,4)}
Utilizing two four-dimensional bits, generally referred to as \emph{quarts}, it is possible to encode up to sixteen different messages in a single ququart. In the time bin encoding frame, a four-dimensional quantum state is a train of four pulses, with specific heights and relative phases. For practical reasons, we chose to encode a limited subset of them, represented by the ones with zero relative phase between every pulse. In this case, by measuring in the \textbf{Z}~basis, Bob gets information about the first quart, while \textbf{X}~basis measurement is necessary for extracting the information contained in the second quart. We perform measurements only on the first quart, since every result achieved on it can be reproduced on the second quart provided that the necessary multi-interferometric setup is implemented. The encoded states, generated as reported in section~\ref{sec:qrac}, are reported in Tab.~\ref{tab:enc_4d}. Fig.~\ref{fig:statesqrac} (bottom) shows the wave function of the produced 4-D 00 state. The measurement bases are a two-dimensional product of the eigenbases of the Pauli operators $\sigma_z$ and $\sigma_x$, as defined in Sec.~\ref{sec:theory}. According to our encoding scheme for $00_{4D}$ state, the probability for the photon of being detected in the first time bin is $a_2^2=75\%$, while the probability of being in one of the last three pulses is always equal to $b_2^2=8.3\%$. These values represent the sizes of the pulses to be set during the state preparation; when we measure in single-photon regime, the first pulse has the correct size, while the clicking probabilities measured for the last three pulses are 4.0\%, 7.7\% and 13.1\% respectively. Although they are slightly different from the desired ones, considering that only a click in the first pulse represents a successful $00_{4D}$ measurement in the time bases, the small variance of the other pulses from the optimal values does not constitute an important issue. However, we note that for the phase-measurement it would correspond to a drop in performance to $72.8\%$, estimated by considering all possible permutations of the generated amplitudes in conjunction with the phases. This contributes to the average performance of $73.95\%$, which would still be above the classical limit of $62.5\%$, should we implement such measurement in the QRAC(2,4) case.


\begin{table}[]
    \centering
    \begin{tabular}{c|c|c|c|c|c}
        Message & Binary encoding & $\ket{00}$ & $\ket{01}$ & $\ket{10}$ & $\ket{11}$ \\
        \hline
         00 & 00 00 & $a_2$ & $+b_2$ & $+b_2$ & $+b_2$ \\
         10 & 01 00 & $b_2$ & $+a_2$ & $+b_2$ & $+b_2$ \\
         20 & 10 00 &$b_2$ & $+b_2$ & $+a_2$ & $+b_2$ \\
         30 & 11 00 & $b_2$ & $+b_2$ & $+b_2$ & $+a_2$ \\
    \end{tabular}
    \caption{Scheme of the encoded states in QRAC(2,4). The coefficients in front of the basis vectors, as defined in Sec.~\ref{sec:theory}, are $a_2=\sqrt{3}/2$ and $b_2=1/(2\sqrt{3})$.}
    \label{tab:enc_4d}
\end{table}

\begin{figure}
    \centering
    \includegraphics[width=\columnwidth]{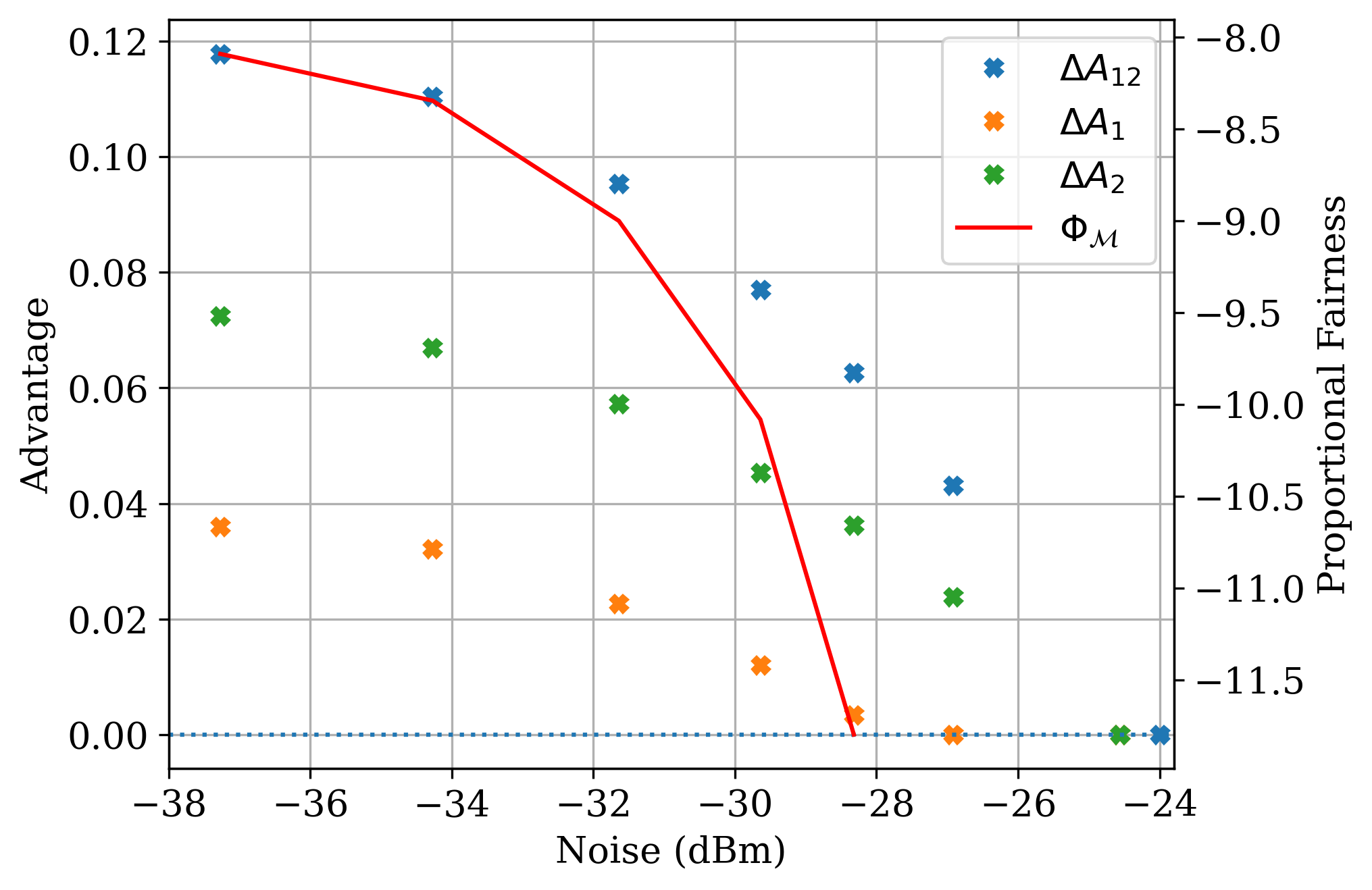}
    \caption{Quantum advantage $\delta A$ of QRAC vs laser power of coexisting classical communication. The red line shows the allocation in terms of proportional fairness $    \Phi_{\mathcal{M}}\left(\mathbf{M}_{E,F}\left[S_{1},S_{2}\right]\right)$ (right axis), as defined in eq.~\ref{eq:allocation1}.}
    \label{fig:advantage}
\end{figure}

As pointed out in Table~\ref{tab:enc_4d}, two quarts are equivalent to four bits. Thus, measuring one of the two quarts is equivalent to extracting the information of two bits out of four. We are also interested in measurements of a subsystem corresponding to only one bit, so we can compare with the case of QRAC(2,2), when one bit (over two) is retrieved. We chose to measure the first bit of the message: in our scheme, this corresponds to measuring if the arrival time of the photon is within the first two time-bins composing the quantum state or within the last two ones. We call $M_1$ the measurement over the subset made by the first and the second time-bin, and $M_2$ the measurement over the two remaining time-bins. The measurement that reports the exact time-bin of arrival, i.e. the measurement of the first two bits, is called $M_{12}$. The results for the unperturbed channel are reported in Tab.~\ref{tab:4D_meas}. Fig.~\ref{fig:advantage} represents the quantum advantage $\delta A$ and the proportional fairness $\Phi_{\mathcal{M}}\left(\mathbf{M}_{E,F}\left[S_{1},S_{2}\right]\right)$ (see Eq.~\ref{eq:allocation1}) achieved emulating ongoing classical transmissions at different optical powers. The lower bounds for the probability \textit{p} of retrieving one or two bits over four are given by the classical $p^{CRAC(2,4)}_{1bit}=75\%$ and $p^{CRAC(2,4)}_{2bits}=62.5\%$, the quantum upper bounds are $p^{QRAC(2,4)}_{1bit}=83.3\%$ and $p^{QRAC(2,4)}_{2bits}=75\%$~\cite{Ambainis_Leung_Mancinska_Ozols_2009}.
\begin{table}[]
    \centering
    \begin{tabular}{c|c|c|c}
         & $M_1$ & $M_2$  & $M_{12}$  \\
         \hline
         \textit{p} & 79.1\% & 82.9\% & 75.1\% \\ 
         $\delta A$ & 0.041 & 0.079 & 0.126 \\
    \end{tabular}
    \caption{Achieved successful probability \textit{p} and quantum advantage $\delta A$ for the three measurement sets in \textbf{Z} basis, QRAC(2,4).}
    \label{tab:4D_meas}
\end{table}

The measured quantum advantage is in accordance with the theoretical $\delta A_{th}=p^{QRAC}-p^{CRAC}$ for the dedicated fiber channel and decreases monotonically when addition of noise appears due to a coexistent classical communication. Therefore, it behaves as a monotone function, as defined in section~\ref{sec:theory}. As a consequence, $\delta A$ can be used as a quantifier of the goodness of the resource allocation between the subsystems~$S_1$ and $S_2$. We saturate the quantifier up to the theoretical threshold, so we can say that our allocation of resources is optimal. We demonstrated that a measurement set made by a pair of product mutually unbiased bases is the most incompatible measurement set, and the followed strategy for state encoding is optimal.

\begin{table}[h]
    \centering
    \begin{tabular}{c|c|c}
         Ref. & $\nu$ (Hz)& Protocol \\
         \hline
         \cite{li2023high} & 2.5 GHz & efficient BB84 \\
         \hline
         \cite{grunenfelder2023fast} & 2.5 GHz & 3 states BB84 \\
         \hline
         \cite{islam2017provably} & 2.5 GHz & HD QKD (4D)\\
         \hline
         \cite{zahidy20224} & 487 MHz & HD QKD (4D)\\
         \hline
         \cite{vagniluca2020efficient} & 297.6 MHz & BB84\\
         \hline
         \cite{mafu2013higher} & 80 MHz & HD QKD \\
         \hline
        \cite{ribezzo2023deploying} & 595 MHz & BB84\\
        \hline
         This work 2D & 298 MHz & QRAC(2,2)\\
         \hline
         This work 4D  & 76 MHz & QRAC(2,4)\\
    \end{tabular}
    \caption{Comparison of the quantum state generation rate of our source with other work employing QKD sources. We chose to not compare the QKD final key rate since in QRAC the final exchanged states rate is directly connected just to the generation rate, the attenuation of the channel and the mean number of photon per pulse. In contrast, in Quantum Key Distribution (QKD), the final key rate is significantly lower than the rate of exchanged quantum states, owing to the extensive post-processing steps involved, including error correction, error verification, and privacy amplification. Because of the similarity of our QRAC setup with systems implementing time-bin encoded QKD protocols, future implementation of our scheme will include improved quantum state generation rate thanks to the insights from the recent QKD schemes.}
    \label{tab:rate_comp}
\end{table}
\section{Discussion}

In this work, we have designed and realized a setup for generating and detecting quantum random access code states, taking special care to provide compatibility of the introduced scheme with the standard telecommunication infrastructure. Taking inspiration from many commercial QKD devices, our QRAC device's design is the first ever realized that employs weak pulses and time-bin encoding instead of real single photon sources.
Our solution offers versatile reconfigurability, enabling us to effectively handle both two- and four-dimensional quantum states. We have exploited the potentialities of the protocol to prove the most recent advances in the theory of the optimal allocation of quantum resources. In particular, our experiment has provided a concrete case of quantum protocols that, exploiting the incompatibility of pair of product MUBs, exhibit advantage over their classical counterparts. Furthermore, we have studied the performance of these protocols by emulating actual field trial conditions and determining thresholds for the proportional distribution of resources in a system that undergoes different levels of quantum noise.

Moreover, our paper presents perspectives for future developments in both the theoretical and experimental aspects. The connection between the resource theory of incompatibility and the QRACs, exploited in this work, suggests a broader relationship with prepare-and-measure protocols. In this regard, it would be interesting to investigate the possibility of a connection between certain aspects of QKD, where MUBs play a crucial role~\cite{Wooters1989, Durt2005, durt2010, Cerf2002, Sheridan2010, Fenerczi2012, Wang2021}, and a potential novel monotone quantifying incompatibility, with an operational interpretation in terms of the key rate, resistance to noise or another aspect of QKD.  On top of the above,  QRACs as a semi-device-independent self-testing tool, have broader implications for advancing quantum distributed computation and communication in a trustworthy and secure manner. Specifically, self-testing offers a distinct form of security compared to secure data exchange by certifying devices in a separate node of the communication network based solely on one-sided information. We remark, that the above remote test aligns with the overarching goals of quantum technology initiatives to promote independent and reliable quantum communication solutions \cite{purohit2023building,EUQuantFlag}.

On the experimental side, our setup provides room for further development. The receiver setup can be integrated with the interferometric stage necessary to read the relative phase of a four-pulse state, to access every bit encoded in a four-dimensional quantum state. If Bob wants to perform a phase measurement of a QRAC(2,4) state, and $\tau$ is the time distance between two following pulses, he needs a first Mach-Zenhder interferometer with a delay line of 2$\tau$, and two additional Mach Zenhder interferometers with a delay line of $\tau$ connected to the two outputs of the first one, with a need of four SPDs in total. This scheme is viable in a four-dimensional Hilbert space, but when scaling up to higher dimensions is required, it encounters non-trivial practical and economic limitations due to the exponential scaling of the number of necessary interferometers.

However, in applications requiring quantum states with higher dimensions than those utilized in this work, a recently introduced scheme allows for reducing the complexity of the receiver setup without necessitating significant technological replacements. By delaying the photons from one output of the first interferometer, it becomes possible to reduce the number of required interferometers and SPDs down to linear scaling with the number of qubits~\cite{ikuta2022scalable}. In future research, we intend to explore states with higher dimensions by incorporating this idea into our receiver module.

In addition, the SPADs can be replaced by Superconducting Nanowire Single Photon Detectors (SNSPDs), which offer outstanding performance in terms of intrinsic dark counts (\mbox{$\tau_{DC}<1\text{ Hz}$}), timing jitter (\mbox{$\sigma_J<15\text{ ps}$}) and detection efficiency ($\eta>90\%$)~\cite{single_quantum_snspd}. It would result in higher transmission rates, stronger resilience to noise, and greater quantum advantage, still closer to the theoretical bounds.
Finally, our transmitter and receiver modules can be designed as Photonic Integrated Circuits (PICs). This would allow for ultra-compact devices while maintaining excellent performance. PIC interferometers are proven to be reliable and stable without the necessity of a phase stabilization stage~\cite{ding2017high}.

In conclusion, the work presented in this paper provides an important starting point for further endeavours aimed at designing and deploying next-generation quantum communication networks. We expect to work towards a fully-fledged field trial in one of the test-bed networks that are currently being rolled out in Europe under the auspices of the EuroQCI initiative of the European Commission~\cite{euroqci}. In the medium term, these networks will transition from simple QKD networks to a \emph{bona fide} Quantum Internet, where quantum resources will be shared between remote nodes for purposes not limited to cryptography. Such networks would certainly require the optimisation of resource allocation, and would therefore benefit from the application of the techniques outlined in this paper.

\begin{acknowledgments}

\vspace{-0.2 cm}

 RS acknowledge the financial support of the Foundation for Polish Science through the TEAM-NET project (contract no. POIR.04.04.00-00-17C1/18-00).
 JCz acknowledges the financial support from NCN DEC-2019/35/O/ST2/01049. RS and JCz also thank partial funding from LASERLAB-EUROPE (grant agreement no. 871124, European Union’s Horizon 2020 research and innovation programme).
 This work was partially supported by the European Union’s Horizon 2020 research and innovation programme, project QUANGO (grant agreement No 101004341) [AX and NF] and project QSNP (grant agreement No 101080116 and No 101114043) [AX]; the European Union's Horizon Europe research and innovation programme, project QUDICE (grant agreement No 101082596) [AX]; and the European Union's Digital Europe Programme, project PRISM (grant agreement No 101111875) [AX and NF].
This work was partially supported by the Center of Excellence SPOC (ref DNRF123), Innovations fonden project Fire-Q (No. 9090-00031B), the NATO Science for Peace and Security program (Grant No. G5485, project SEQUEL), the programme Rita Levi Montalcini QOMUNE (PGR19GKW5T), the EraNET Cofund Initiatives QuantERA within the European Union’s Horizon 2020 research and innovation program grant agreement No.731473 (project SQUARE), the Project EQUO (European QUantum ecOsystems) which is funded by the European Commission in the Digital Europe Programme under the grant agreement No 101091561, the Project SERICS (PE00000014) under the MUR National Recovery and Resilience Plan funded by the European Union - NextGenerationEU, the Project QuONTENT under the “Progetti di Ricerca@CNR” program funded by the Consiglio Nazionale delle Ricerche (CNR) and by the European Union - PON Ricerca e Innovazione 2014-2020 FESR - Project ARS01\_00734 QUANCOM.
\end{acknowledgments}

\bibliographystyle{unsrtnat-AQT}
\bibliography{references}

\end{document}